\documentclass[sigconf,balance=false]{acmart}
\setcopyright{none}
\acmConference[]{}{}{}

\acmYear{2027}
\copyrightyear{2027}
\acmDOI{}
\acmISBN{}
\usepackage{booktabs,amsmath,graphicx,tabularx,xurl}
\usepackage{microtype}
\graphicspath{{figures/}}
\hypersetup{pdftitle={AI and Mathematical Work on Mastodon: Attention, Topics, and Target-Specific Evaluations},pdfauthor={Takayuki Uchiba, Kunihiro Miyazaki}}
\title{AI and Mathematical Work on Mastodon: Attention, Topics, and Target-Specific Evaluations}
\author{Takayuki Uchiba}
\affiliation{
  \institution{Sugakubunka Co., Ltd.}
  \city{Tokyo}
  \country{Japan}
}
\affiliation{
  \institution{ZEN Mathematics Center}
  \country{Japan}
}
\email{takayuki.uchiba@sugakubunka.com}

\author{Kunihiro Miyazaki}
\affiliation{
  \institution{The University of Tokyo}
  \city{Tokyo}
  \country{Japan}
}
\email{kunihirom@acm.org}

\begin{document}
\begin{abstract}
How does a mathematical community discuss AI on a federated social platform? We study 24,920 public posts from January--September 2026 by 272 accounts in a fixed roster of 1,002 Mastodon accounts whose profiles indicate mathematical research or study. LLM-assisted extraction identifies 1,191 AI-for-mathematics posts by 101 accounts. Their share of all posts rises from 1.1\% in January to 11.5\% in September, with contributing accounts increasing from 10 to 59. We represent discussion through linked target--aspect--stance triples, separating what is discussed from the property assessed and the author's evaluation. From July to September, capability and usefulness decline as a share of related discussion, while social and institutional concerns increase; changing contributors account for much of the latter shift. Among 38 accounts evaluating both technical quality or utility and communication, credit, or verification procedures, the procedural negative-component fraction is 19.4 percentage points higher on average (95\% account-bootstrap interval: 3.8--35.1). Matching targets within posts gives a similar direction. Repeated extraction varies substantially at the level of individual triples, although replacing the repeated sample preserves the aggregate contrast. Around OpenAI's October 6 mathematics release, related-post share rises from 9.6\% to 13.6\%, and contributing accounts increase from 13 to 28. Post-release discussion includes appreciation of the results' potential mathematical value and applications alongside criticism of promotion centered on result counts, limited methodological disclosure, and verification arrangements.
\end{abstract}
\ccsdesc[500]{Human-centered computing~Social media}
\ccsdesc[300]{Information systems~Sentiment analysis}
\keywords{AI for mathematics, Mastodon, federated social media, target stance, aspect-based analysis, computational social science}
\maketitle

\section{Introduction}
Federated social platforms host professional discussion in which research artifacts, commentary, and personal assessments appear together. Measuring opinion in these streams requires distinguishing the object being discussed from the property assessed, separating sharing from endorsement, and accounting for changes in who contributes. We address this Web-related measurement problem through public mathematical discussion on Mastodon.

AI can assist mathematical work through computation, explanation, formalization, and proof production. These uses raise different evaluative questions. A researcher may find a system useful while criticizing a paper's attribution or the arrangements for verifying a public claim. A positive response to a declaration criticizing AI is approval of that declaration; it need not be approval of AI. Treating such expressions as a single attitude loses the relationship between an evaluation and its object.

We analyze posts from 1,002 accounts selected through profile evidence of mathematical research or current study. The historical corpus contains 24,920 public posts from January--September 2026. Rather than assign each post one overall evaluation, we extract zero or more linked \emph{target--aspect--stance} triples. A target is an identifiable object, an aspect is a property or procedure concerning it, and stance is the posting author's evaluation of that pair. This representation supports both descriptions of discussion and comparisons of evaluations within accounts. A separate October collection supports a temporal comparison of attention, participation, and positive and negative expressions on each aspect around a mathematical release.

Our research questions are:
\begin{enumerate}
\item[\textbf{RQ1.}] How do attention to AI for mathematics and participation in that discussion change from January to September?
\item[\textbf{RQ2.}] Which targets and aspects are discussed, and how does their distribution change over time?
\item[\textbf{RQ3.}] Within the same accounts, how do evaluations of technical quality and utility differ from evaluations of communication, credit, and verification procedures?
\item[\textbf{RQ4.}] How do attention, participation, and aspect-specific positive and negative expressions change around OpenAI's October 6 mathematics release?
\end{enumerate}

We find increasing attention, a relative shift toward social and institutional aspects, and more negative procedural evaluations within accounts. The shift toward social and institutional aspects is partly compositional. Our contribution is an empirical account of these distinctions and an auditable measurement procedure connecting individual expressions to aggregate patterns.

\section{Related Work}
Research on scientists' social-media participation illustrates both the value of professional communities as study populations and the difficulty of identifying them from public traces~\cite{ke2017scientists}. Mastodon's decentralized organization makes it a distinct setting for studying online communities~\cite{zignani2018mastodon}. Miyazaki et al.\ find that perceptions of generative AI vary with occupation and use~\cite{miyazaki2024}. We examine variation within one mathematical community, including changes in the accounts contributing to each period. This distinction between observed traces and the population or behavior they represent follows the measurement and representation concerns articulated by Sen et al.~\cite{sen2021tedon}.

Stance detection distinguishes a position toward a target from the sentiment of an entire text~\cite{sobhani2016stance}. Aspect-based sentiment analysis further separates the object, its assessed aspect, and polarity. Wan et al.\ explicitly model target--aspect--sentiment triples~\cite{wan2020target}; Zhang et al.\ generate aspect-category, aspect-term, opinion-term, and sentiment quadruples~\cite{zhang2021quad}. Our representation follows this separation while retaining exact evidence, attribution, conditionality, and mentions without expressed evaluation. It is therefore related to these tasks, but its six stance states and source-bounded interpretation are tailored to mathematical discussion.

LLMs have been used for joint aspect extraction and polarity classification~\cite{simmering2023absa}, while broader evaluations find that structured sentiment tasks remain challenging~\cite{zhang2024reality}. Work on computational social science emphasizes task-specific assessment of LLM annotation~\cite{ziems2024css,pangakis2023validation}. We examine variation through fresh extraction runs and downstream sensitivity, distinguishing the stability of individual annotations from the stability of a particular aggregate comparison.

\section{Data and Methods}
\subsection{Account roster and collection}
We constructed a roster from the opt-in profile directories of \nolinkurl{mathstodon.xyz} and \nolinkurl{mastodon.social}~\cite{mastodondirectory,mastodondirectoryapi}. Two directory traversals on October 1--2, 2026 recorded 679,126 distinct accounts: 4,407 and 674,719, respectively. After rule-based filtering, GPT-6 Astra assessed saved biographies and profile fields for explicit evidence of mathematical research, a research position, or current study of mathematics. Interest or a past degree alone did not qualify; 469 ambiguous profiles were excluded. Identical profile inputs reused eligibility decisions. Posting activity and views about AI were not selection criteria.

The resulting roster contains 1,002 accounts: 690 on Mathstodon and 312 on \nolinkurl{mastodon.social}. Profiles indicated researchers (711), students (279), or both (12). Accounts, rather than inferred individuals, are the units of analysis.

On October 3, we retrieved available public posts dated January 1--September 30, 2026 (UTC), including replies and excluding boosts. We paginated the account-status endpoint~\cite{mastodonaccounts,mastodonpagination} until it returned no further history. The corpus contains 24,920 posts from 272 accounts; 730 roster accounts had no retrieved post in that window. A separate collection on October 14--15, with a September 29 backfill completed on October 16, covers September 29--October 13. It contains 1,437 observations from 116 accounts. The two collections share 199 original posts and are analyzed separately. Collection details are in Appendix~\ref{app:collection}.

\begin{table}[tp]
\caption{Corpus and extraction coverage. Account counts are distinct within each row. Complete extraction includes related posts with no identifiable target.}
\label{tab:flow}
\centering
\begin{tabular}{lrr}
\toprule & Accounts & Posts\\
\midrule
\multicolumn{3}{l}{\emph{January--September historical corpus}}\\
Retrieved public posts & 272 & 24,920\\
AI4Math-related & 101 & 1,191\\
Related, complete extraction & 99 & 1,160\\
Related, complete with triples & 99 & 1,150\\
Uncertain relevance & 67 & 312\\
Source text unavailable & 11 & 22\\
\midrule
\multicolumn{3}{l}{\emph{September 29--October 13 collection}}\\
Retrieved public posts & 116 & 1,437\\
AI4Math-related & 33 & 167\\
Related, complete extraction & 33 & 161\\
Uncertain relevance & 27 & 76\\
Source text unavailable & 1 & 2\\
\bottomrule
\end{tabular}
\end{table}

\subsection{Source-bounded triple extraction}\label{sec:extraction}
All annotation stages used GPT-6 Astra LLM agents. An input contains the post body, content warning, available media descriptions, visible URLs, and at most two archived ancestors. Ancestors created or edited after the focal post were omitted. Account identifiers and timestamp metadata were withheld, although references within the text remained visible. Agents did not open links or retrieve later replies.

The first stage processed all 26,357 collection-specific observations, assigning relevance and extracting target--aspect candidates in separate fields. AI4Math relevance requires a connection between AI and mathematical reasoning, research, education, or professional work. Formalization alone is insufficient, as is a mathematical paper attributed to an AI company without a recoverable connection to AI use. Comprehensible mathematical resources connected to AI only by a hashtag are marked \emph{tag-only}; opaque references remain uncertain. Unavailable source text is recorded separately.

A second stage assigned stance using the original input and candidates. A fresh third-stage agent reviewed the source with the draft visible, allowing additions, removals, splits, and corrections with reasons. These stages covered all 1,786 observations initially marked related or uncertain, or containing a candidate or unresolved reference. The other 24,571 were unrelated with empty candidate and unresolved-reference sets. Review could also amend relevance. The finalized outputs contain 3,033 triples across both collections and relevance categories.

Stance takes six values: \emph{positive}, \emph{negative}, \emph{explicitly neutral}, \emph{mixed}, \emph{no expressed stance}, and \emph{unclear}. Sharing, a question, or an unendorsed quotation does not by itself express stance. Mixed requires opposing evaluations of the same target and aspect. Different aspects form separate triples; qualifications such as conditionality and tentativeness are retained. Exact source quotations support target identification, aspects, and expressed evaluations. An ancestor or URL can resolve a target but cannot alone establish the focal author's endorsement.

\subsection{Target and aspect categories}\label{sec:categories}
We developed the category system on 135 historical observations classified as related during preliminary screening: ten development and five pilot observations per month. Both subsets informed the codebook and remain part of the corpus. The codebook was frozen before production mapping and aggregate-outcome inspection.

Mapping agents received the source and finalized target--aspect pairs, without stance labels or aggregate results. They assigned target type first and aspect second, preserving actual pairs rather than forming all combinations. Table~\ref{tab:taxonomy} defines the substantive categories. A paper remains a mathematical or technical work when assessed for correctness, exposition, or attribution; the aspect does not change its object type. Raw descriptions and a 36-node auxiliary hierarchy preserve detail beneath the main aspects.

\begin{table*}[tp]
\caption{Substantive target and aspect categories. Target types also include other and unresolved; aspects additionally include other, general evaluation, unspecified, and unresolved. These residual states are retained rather than forced into a substantive family.}
\label{tab:taxonomy}
\centering
\begin{tabularx}{\textwidth}{lX}
\toprule
Category & Definition\\
\midrule
\multicolumn{2}{l}{\emph{Target: what is discussed}}\\
Computational systems/resources & Models, software, proof assistants, libraries, databases, and technical ecosystems used as resources.\\
Mathematical/technical works & Particular proofs, answers, papers, preprints, formalizations, or other produced artifacts.\\
Statements/commentary/media & Announcements, declarations, commentary, reviews, talks, and media items as communicative objects.\\
People/organizations & Individuals, groups, companies, and institutions as actors.\\
Activities/workflows/arrangements & Research or educational activities, collaborations, adoption practices, and institutional arrangements.\\
\midrule
\multicolumn{2}{l}{\emph{Aspect: which property or procedure is discussed}}\\
Capability/usefulness & Capacity or practical usefulness for mathematical, supporting, or educational tasks.\\
Correctness/technical validation & Mathematical or computational correctness, technical checking, and validity of capability measurements.\\
Understanding/contribution & Reasoning, creativity, originality, importance, and mathematical content or structure.\\
Usability/work required & Intelligibility, operational usability, speed, integration, and the human effort required.\\
Reporting/disclosure/attribution & Representation of work, access to evidence, provenance, credit, and responsibility in reporting.\\
Review/verification/release & Organization, requirements, responsibilities, and burdens of reviewing, checking, endorsing, or releasing work.\\
Social/professional/institutional & Professional consequences, norms, commercial power, fairness, access, and research experience.\\
Costs/environment & Monetary or material costs and ecological effects.\\
\bottomrule
\end{tabularx}
\end{table*}

For RQ3, the first four substantive aspects form \emph{technical quality/utility}; reporting/disclosure/attribution and review/verification/release form \emph{communication/credit/verification}, abbreviated \emph{procedural}. Social and institutional relations and costs remain separate. These are aspect families: a tool, paper, statement, or activity can be evaluated on either family.

\subsection{Aggregation and comparisons}\label{sec:aggregation}
RQ1 counts all 1,191 posts classified as AI4Math-related because measuring attention requires a relevance decision, not a complete set of triples. This includes 31 posts whose target--aspect extraction is partial or unresolved. RQ2--3 require complete extraction and therefore use 1,160 related posts, of which 1,150 contain 2,506 triples (Table~\ref{tab:flow}). An absent target contributes no evaluation; complete posts without a target remain in the denominator for aspect-mention shares.

The cross-classification collapses repeated triples to one \emph{post--target-category--aspect-category} unit. Target, aspect, and family margins are each deduplicated again at their own level. For a unit $u$, the expressed-evaluation indicator $E_u$ is one when at least one represented triple has a positive, negative, explicitly neutral, or mixed stance, and zero otherwise. Thus, $E_u$ identifies units included in the evaluation denominator; it does not indicate a particular polarity. The positive-component indicator $P_u$ is one if any triple is positive or mixed, and the negative-component indicator $N_u$ is one if any triple is negative or mixed; both are zero otherwise. An explicitly neutral evaluation contributes to the denominator but to neither component. Units containing only no expressed stance or unclear do not enter this denominator. A broad unit may contain both components without any raw triple being mixed.

For account $a$ and aspect family $f$, define the positive- and negative-component fractions and the RQ3 contrast as
\begin{equation}
\begin{aligned}
r^{+}_{af}&=\frac{\sum_{u\in U_{af}}P_u}{\sum_{u\in U_{af}}E_u},\qquad
r^{-}_{af}=\frac{\sum_{u\in U_{af}}N_u}{\sum_{u\in U_{af}}E_u},\\
\Delta&=\frac{1}{|B|}\sum_{a\in B}
\bigl(r^{-}_{a,\mathrm{proc}}-r^{-}_{a,\mathrm{tech}}\bigr).
\end{aligned}
\label{eq:paired}
\end{equation}
Here $U_{af}$ contains deduplicated post--family units; $\mathrm{proc}$ and $\mathrm{tech}$ denote procedural and technical aspects. Fractions are defined only for accounts with a nonzero evaluation denominator, and $B$ contains accounts evaluating both families. We average each component fraction equally over evaluating accounts; RQ3 compares the negative components. The same fractions are calculated for target--aspect cells using their own deduplicated units. We also report pooled post rates and rates among all mentions. Mixed evaluations contribute to both components, so positive and negative fractions need not sum to one.

Monthly aspect-mention rates divide each account's mentioning posts by its complete related posts and average over all complete-related contributors, including zero mentions. RQ2 also reports pooled related-post and all-post shares. To distinguish changes within contributors from composition, we decompose July--September changes into continuing-account rates, continuing-account weights, entry, and exit (Appendix~\ref{app:decomposition}). Entry and exit mean contributing in only one comparison month. All monthly series are reported; this contrast is exploratory.

\begin{figure*}[t]
\centering\includegraphics[width=\textwidth]{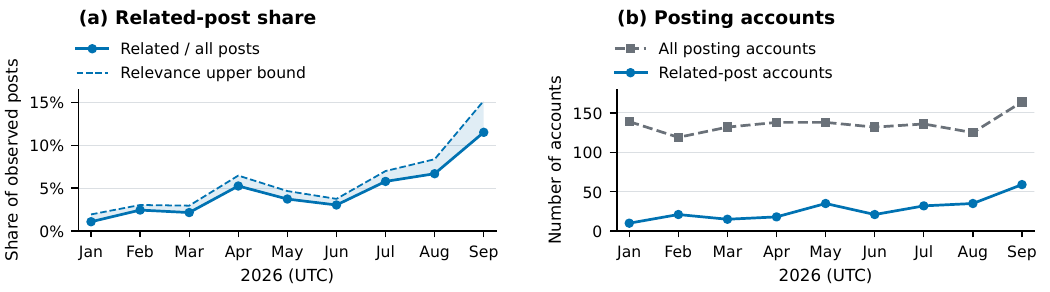}
\caption{Historical attention and participation. Left: related posts divided by all observed monthly posts; the upper dashed boundary also includes uncertain relevance and unavailable text. Shading represents an allocation bound, not a confidence interval. Right: accounts contributing any post and accounts contributing a related post. Dates use UTC.}
\Description{Two panels show monthly related-post shares and account counts. Related shares rise from 1.09 percent in January to 11.49 percent in September with intervening fluctuations; related contributors rise from 10 to 59.}
\label{fig:attention}
\end{figure*}

For RQ3, we additionally restrict the comparison to posts that evaluate both aspect families for the same explicitly identified target. We calculate the procedural-minus-technical difference for each post--target unit, average these differences within account, and then average equally across accounts. A stricter cross-post match uses only explicit source URLs or system names. Five separate sensitivity analyses retain the January--September window and change one rule at a time: exclusion of tag-only relevance; retention of unqualified evaluations only; counting negative components only when no positive component is present; inclusion of resolved units from incompletely extracted related posts; and inclusion of resolved units from uncertain-relevance posts with complete extraction. RQ3 intervals use 5,000 paired-account bootstrap draws, with annotations and the random seed fixed. Signs are determined by integer or rational arithmetic; intervals are not estimated for fewer than two accounts.

\subsection{Temporal comparison and measurement assessment}\label{sec:assessment}
RQ4 compares September 29--October 5 with October 6--12 around OpenAI's October 6 release~\cite{openai2026mathematicsrelease}, treating the UTC release date as the first post-release day. That day's aggregate can include posts preceding the announcement; a sensitivity analysis excludes it. Attention is related posts divided by all posts; participation is related contributors divided by the fixed roster of 1,002. Evaluation analyses use all complete AI4Math-related posts in these windows and all eight substantive aspects separately, including posts unrelated to OpenAI. Neither release attribution nor paired-account membership selects this corpus.

For each post--aspect unit, positive and negative components are Boolean unions of its triples; mixed contributes to both. We distinguish two daily rates for each component. \emph{Expression incidence} divides an account's component-bearing posts by all its complete related posts that day, then averages equally over that day's complete-related contributors, including zero mentions of the aspect. \emph{Conditional stance} instead divides by that account's evaluated posts for the aspect and averages only accounts with a nonzero evaluation denominator. No evaluation yields zero incidence but undefined conditional stance. Incidence measures the presence of an evaluation in related discussion; conditional stance measures its share among expressed evaluations. Residual categories remain in supplementary tables.

With $x_t$ the UTC calendar-day distance from October 6 and $D_t=I(x_t\geq 0)$, we fit each outcome using a regression discontinuity in time (RDiT) specification~\cite{hausman2018rdit}:
\begin{equation}
y_t=\alpha+\beta x_t+\tau D_t+\gamma x_tD_t+\varepsilon_t.
\label{eq:temporal}
\end{equation}
Dates have equal weight. The cutoff gap $\tau$ compares the two fitted lines at $x_t=0$; $\gamma$ is their slope difference. All sixteen primary aspect--polarity incidence series have seven observed dates per side. For the additional conditional-rate fits, our reporting rule requires five defined dates per side and a rank-four design; undefined dates retain their calendar spacing. We treat the fitted gaps as descriptive and report no $p$-values or confidence intervals. We retain out-of-range linear predictions and report their diagnostics. The cutoff coincides with a statement by the Advisory Group on Mathematics and Artificial Intelligence (AGMAI)~\cite{agmai2026oct6}, so it does not isolate OpenAI's effect. Appendix~\ref{app:event} reports counts, conditional-rate support, and sensitivity to five-/six-day windows, relevance rules, pooled-post weighting, and release-day omission.

The authors checked the aggregate findings for plausibility; no independently human-coded accuracy benchmark was used. Repeatability and omission checks are described in Appendix~\ref{app:execution}.

\begin{figure*}[t]
\centering\includegraphics[width=\textwidth]{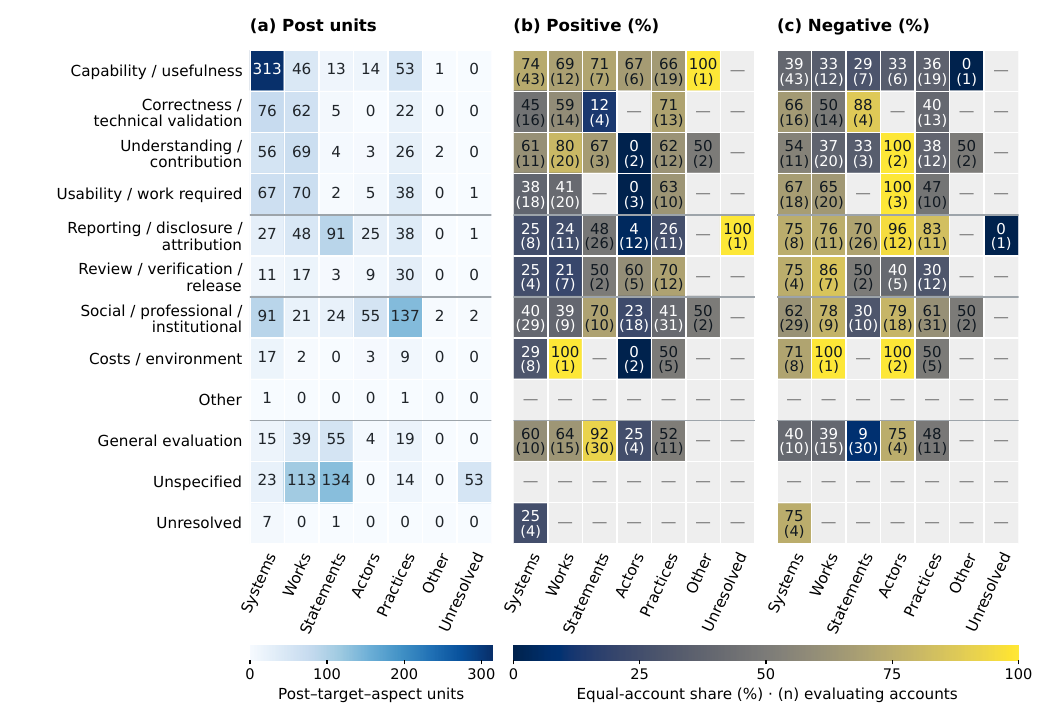}
\caption{Target--aspect cross-classification. Left: distinct mentioning posts. Center and right: equal-account positive- and negative-component percentages among expressed evaluations, on the same 0--100\% scale; parentheses give evaluating-account counts. A dash means no expressed evaluation. Mixed contributes to both components. Target labels abbreviate Table~\ref{tab:taxonomy}; a post can enter several cells.}
\Description{Three heatmaps cross twelve aspect categories with seven target categories. The count panel has 313 system-capability mentions, 137 activity-social mentions, and 91 statement-reporting mentions. The other panels separately show positive- and negative-component percentages, evaluating-account counts, and undefined cells.}
\label{fig:cross}
\end{figure*}

\section{Results}
\subsection{RQ1: Attention and participation}
AI4Math-related posts comprise 4.78\% of the historical corpus (1,191/24,920). Their share rises, non-monotonically, from 1.09\% in January (28/2,580) to 11.49\% in September (357/3,106); related-post contributors increase from 10 to 59 (Figure~\ref{fig:attention}). Assigning every uncertain or unavailable observation to related gives upper bounds of 1.94\% and 15.16\%, respectively. Thus the ordering of these two months persists under these extreme relevance allocations.

January examples included AI-assisted mathematical visualizations and a chatbot-generated theorem-dependency diagram, reported as correct but requiring manual rearrangement of overlapping elements. September posts discussed proof formalization and proposals for community participation in AI research governance. These examples illustrate the practical work and research organization encompassed by the attention measure.

\subsection{RQ2: Targets, aspects, and changing contributors}
The most frequent target--aspect combination is computational systems/resources with capability/usefulness (313 posts). Activities and arrangements with social/institutional aspects occur in 137 posts, and statements/commentary with reporting/disclosure/attribution in 91. Figure~\ref{fig:cross} shows mentions and both positive and negative evaluations, preserving non-evaluative sharing: statements and research works often have an unspecified aspect, while evaluative cells have smaller supports. Technical and procedural aspects occur within several target types.

\begin{figure*}[t]
\centering\includegraphics[width=\textwidth]{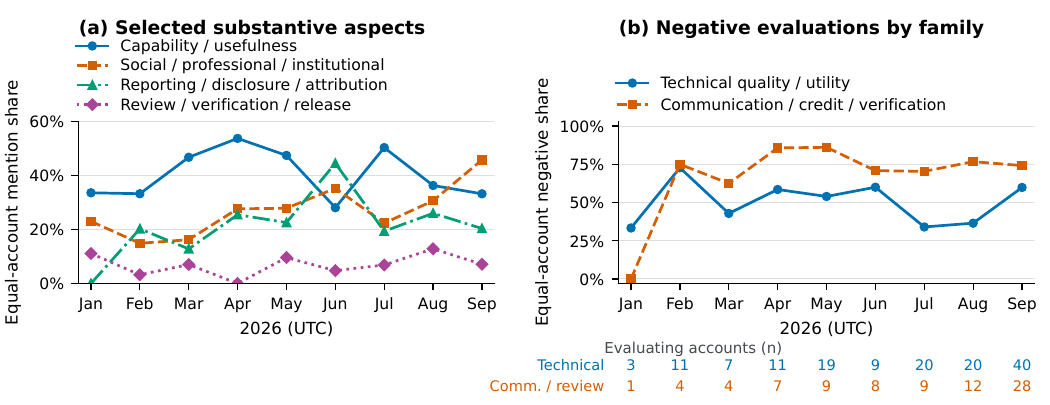}
\caption{Monthly aspect mentions and evaluations, with equal account weights. Left: four substantive aspect-mention rates, averaged across complete-related contributors including zero mentions. Right: negative-component rates among expressed evaluations in each family; evaluating-account counts appear below the axis. Contributor sets vary by month and, on the right, by family.}
\Description{Capability mentions fall in relative share from July to September, while social and institutional mentions increase. Procedural negative-component rates exceed technical rates from February through September. January has one procedural evaluating account.}
\label{fig:monthly}
\end{figure*}

From July to September, capability/usefulness falls from 48.1\% to 28.4\% of complete related posts (75/156 to 99/349), and from 50.3\% to 33.2\% with equal account weights. Its count nevertheless rises, as does its share of all posts, from 2.66\% to 3.19\%. The change is therefore a decline in relative emphasis within related discussion, not a decline in observed capability mentions. July examples ranged from reported help with generalized cohomology calculations using techniques not specified in the prompt and discussion of porting formalizations between proof assistants to frustration with errors in reproducing Riemann's argument for analytic continuation of the zeta function. September posts described automation of mechanical steps in mathematical visualization and a model's initial explanation of binomial-coefficient patterns that provided a starting point for a fuller argument.

Social/professional/institutional aspects increase from 26.3\% to 33.0\% of complete related posts, or 22.4\% to 45.8\% with equal account weights. September posts give this increase concrete content through concerns about corporate influence and proposals for wider model access and community governance. Reporting/disclosure/attribution rises from 16.7\% to 23.5\% by posts but only 19.4\% to 20.4\% by accounts; one post asked how many unsuccessful problem-solving attempts lay behind reported successes, while another questioned credit for prior work. Review/verification/release changes little (5.8\% to 6.0\% by posts; 6.9\% to 7.1\% by accounts). Its content ranges from welcoming human review of a formalization in July to questioning unpaid verification burdens and proposing collaborative exposition in September. The different trajectories argue against combining all social and procedural concerns into one topic (Figure~\ref{fig:monthly}).

September reactions also concerned OpenAI's September 8 release of a claimed AI-generated Navier--Stokes proof, accompanied by a paper and Lean formalization~\cite{openai2026navierstokes}. One post praised the technical achievement while recognizing prior conceptual contributions by humans. Another endorsed criticism that credit focused on the final breakthrough rather than earlier contributors. A third distinguished logical validity from mathematical understanding and emphasized the unpaid work of reviewing the proof.

There are 19 complete-related contributors in both July and September, 40 only in September, and 12 only in July. The 23.4-point equal-account increase in social/professional/institutional mentions decomposes into +4.3 points from continuing-account rate changes, $-10.2$ from their changed weights, +33.1 from September-only contributors, and $-3.8$ from July-only contributors. Much of the overall shift therefore reflects who contributes, rather than a uniform change among continuing accounts. One September-only contributor advocated wider model access and a pilot implementation of community recommendations, illustrating that the added discussion included constructive proposals as well as criticism.

\subsection{RQ3: Evaluations within accounts}
The procedural family has 251 mentioning posts and 171 evaluated posts from 46 evaluating accounts; the technical family has 642 mentions and 331 evaluations from 71 accounts. Equal-account negative-component rates are 74.6\% and 47.7\%, respectively; pooled post rates are 76.0\% and 55.3\%. These marginal comparisons use different account sets.

Among the 38 accounts evaluating both families, the mean procedural-minus-technical difference is +19.4 percentage points (95\% interval [3.8, 35.1]); 19 differences are positive, 11 zero, and eight negative (Table~\ref{tab:paired}). Restricting the comparison to posts that evaluate both aspect families for the same target leaves 27 post--target units across 14 accounts and a mean difference of +15.5 points [2.4, 32.1]. Four account-level differences are positive and ten zero. Cross-post matching on explicit target identifiers leaves only one account, so it does not support a broader matched-target comparison.

In one matched post, the author could check the matrix encoding in generated code against known mathematical sources but could not identify where the matrices originated. Another asked how correct AI-generated proofs could become publishable when their authors lacked the ability or willingness to prepare them. Both contrasted technical and procedural aspects of the same target within the same post.

\begin{table}[tp]
\caption{Procedural-minus-technical negative-component differences (percentage points). $A$ counts paired accounts; signs count positive/zero/negative account differences. Intervals resample paired accounts.}
\label{tab:paired}
\centering
\begin{tabularx}{\columnwidth}{Xrrl}
\toprule Comparison & $A$ & Mean [95\% interval] & Signs\\
\midrule
Within account & 38 & 19.4 [3.8, 35.1] & 19/11/8\\
Same post and target & 14 & 15.5 [2.4, 32.1] & 4/10/0\\
\bottomrule
\end{tabularx}
\end{table}

Under each of the five sensitivity conditions described in Section~\ref{sec:aggregation}, the procedural-minus-technical difference remains positive, with its 95\% account-bootstrap interval above zero (Appendix~\ref{app:sensitivity}).

Procedural rates exceed technical rates in February--September, but not January, whose procedural denominator is one evaluated post. Technical negative-component rates rise from 34.0\% in July to 59.8\% in September. These monthly descriptions use changing contributor sets; they are not paired changes over time.

\begin{figure*}[t]
\centering\includegraphics[width=\textwidth]{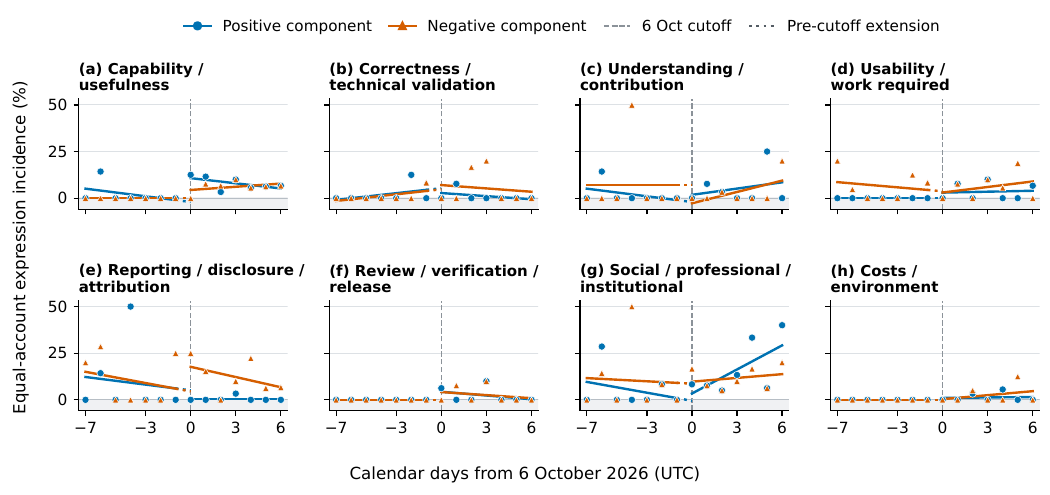}
\caption{Daily positive and negative expression incidence for each substantive aspect. Each rate averages account-specific component-post fractions over all complete-related contributors that day, including zero mentions. Mixed can contribute to both components. Points are observations, with October 6 included in the post period. Solid lines are separate pre/post linear fits; dotted extensions carry the pre-period lines to the cutoff. Linear predictions outside the percentage range are retained. All dates are UTC.}
\Description{Eight panels separately display positive and negative expression incidence. Main fitted gaps are minus 4.63 and plus 13.20 points for reporting, disclosure, and attribution, and plus 4.33 and plus 4.18 points for review, verification, and release.}
\label{fig:aspecttemporal}
\end{figure*}

\subsection{RQ4: Aspect-specific changes around the October release}
Between the pre- and post-release weeks, related posts increase from 56/583 (9.6\%) to 103/757 (13.6\%), and related contributors from 13 to 28. Daily segmented fits give cutoff gaps of +5.31 percentage points for related-post share and +0.91 points for roster participation. The aspect analysis uses 54 complete related posts in the pre period and 99 in the post period.

Figure~\ref{fig:aspecttemporal} separates positive and negative expression incidence for each aspect. Reporting/disclosure/attribution has fitted gaps of $-4.63$ points for positive and $+13.20$ for negative expressions. Its mean daily negative incidence rises from 10.5\% to 12.2\%, while positive incidence falls from 9.2\% to 0.5\%. Review/verification/release has gaps of $+4.33$ and $+4.18$ points, respectively: both polarities appear after a pre-period with no evaluated posts for this aspect. The post-period contains two positive and two negative review posts. Table~\ref{tab:rq4aspects} reports every aspect, both polarities, and their counts.

Positive reporting posts before October 6 endorsed sharing proof artifacts or conditionally welcomed companies sharing solved-problem material. Afterwards, examples criticized sparse disclosure of prompting, model choice, and problem selection, OpenAI's use of headline numbers in promotion, and sensational media framing. One compared the few reasoning-trace summaries supplied with the much larger announced problem count. Other posts anticipated useful applications of OpenAI's results or defended their potential mathematical value even when derived from existing knowledge. Review-related criticism concerned checking released texts and the burden of verifying unexamined proofs. One argued that checking plausible but unexamined proofs would divert mathematicians from their own research. One positive review post defended OpenAI's publication approach, citing repository restrictions, while criticizing its broader conduct; another endorsed checking an AI-assisted LaTeX conversion through its diff.

The reporting result depends on the specification. Its negative gap is $+10.52$ points with six days per side, but $-3.29$ with five; excluding tag-only relevance gives $-1.93$ in the seven-day window. Its positive gap also changes sign in shorter windows. Review's positive and negative incidence gaps remain positive across the specified windows, relevance rules, weighting choices, and release-day omission, but rest on four evaluated post--aspect units in the main comparison.

All sixteen aspect-specific positive and negative incidence fits are estimated from the full pre/post AI4Math corpus. They measure the frequency of each expression among complete related posts. The complementary analysis conditions on an expressed evaluation of the aspect; its smaller daily denominators and the reporting rule for those additional fits are detailed in Appendix~\ref{app:event}.

\subsection{Extraction repeatability}
Across pairs of the three pre-review runs on 180 observations, relevance agrees on 88.3--91.7\%, whereas exact target-category--aspect-category--stance set agreement is 38.9--47.2\%; mean set Jaccard similarity is 52.2--59.2\%. Triple counts are 299, 325, and 337. Fine-grained extraction is thus variable even when relevance agrees.

Replacing only these sampled observations in the finalized main data with each repeat yields paired differences of +21.5 points [6.8, 35.4] and +23.7 [7.8, 38.0], with 37 paired accounts each. This limited perturbation preserves the contrast while leaving unsampled annotations fixed. The 360-observation omission check produces no relevance changes, candidates, or unresolved references. These are separate results from the draft-visible source review (Appendix~\ref{app:execution}).

\section{Discussion}
\paragraph{Implications for mathematical work.}
RQ2 and RQ3 show continuing interest in AI's practical uses alongside demands for explanation, attribution, and verification. Capability/usefulness mentions increase in number despite their declining share of related discussion. Within accounts evaluating both families, procedural aspects attract higher negative-component fractions than technical aspects on average, with the same direction when comparing the same target within one post. These results distinguish the usefulness of outputs from evaluations of how they are presented and checked.

RQ4 shows this distinction around a release: interest in mathematical value and applications coexists with criticism of promotion centered on result counts, limited disclosure, and verification burdens. These findings suggest that releases should state how results were obtained, their relation to prior work, and what has been verified. Alongside useful mathematical outputs, developers and research communities need information and procedures that enable others to understand the results, identify their sources, and check their validity.

\paragraph{Analytical interpretation.}
RQ2's growth in social and institutional discussion largely reflects changing contributors, so it does not establish a uniform shift in participants' interests. RQ3's within-account comparison shows that the technical--procedural contrast is not solely a difference between contributor sets.

In RQ4, criticism of promotion and disclosure is observable, but the negative-expression incidence gap at the cutoff for reporting/disclosure/attribution changes sign across windows and relevance rules. An increase for this aspect is therefore not robust to those choices. Replacing the repeated-extraction sample preserves the RQ3 contrast, although individual triples vary.

\section{Scope and Ethics}
The opt-in directories of two servers omit unlisted accounts and other servers. Contemporary profiles do not reconstruct earlier membership; accounts are not necessarily distinct people. Exhaustive pagination retrieves available history but cannot recover deleted or restricted posts. Missing ancestors and edits limit contextual interpretation. We observe public expression in a selected, retained corpus, not population attitudes, exposure, diffusion, or causal effects of announcements. Without comparisons with X, Bluesky, or MathOverflow, we cannot determine which patterns are specific to Mastodon.

GPT-6 Astra was used for profile assessment and post annotation. The paper reports aggregates without handles or verbatim user examples. Record-level evidence and re-linkable identifiers are retained in the author analysis archive, separate from the aggregate manuscript source. 

We also used GPT-6 Astra to assist with coding for data analysis and manuscript drafting.

\section{Conclusion}
AI4Math attracts increasing attention in this mathematical community's Mastodon discussion, with more contributing accounts and a July--September shift toward social and institutional aspects. Changing contributors account for much of this shift. Among accounts evaluating both technical and procedural aspects, negative-component fractions are higher on average for communication, credit, and verification procedures than for technical quality and utility. Discussion around OpenAI's October release combines interest in potential mathematical value and applications with criticism of promotion centered on result counts and limited methodological disclosure. These findings show how expectations of useful mathematical outputs can coexist with concerns about how research is presented and checked. For AI4Math developers and research communities, they motivate tying promotional claims to inspectable evidence, making contributions traceable, and providing feasible verification arrangements.

\label{end:main}
% References continue after the main text; no forced page break.
% Check the final PDF: main content and its figures must end by page 8.
\label{start:references}
\bibliographystyle{ACM-Reference-Format}
\bibliography{references}
\appendix

\section{Collection and Source Context}\label{app:collection}
Directory discovery ran from October 1, 15:17 to October 2, 16:08 UTC. The combined observations span changing directory membership, not a simultaneous census. Historical retrieval used 7,693 requests: 1,002 account lookups and 6,691 post requests. Fifteen transport failures were followed by successful pagination to the endpoint's end. Raw final response bodies are not all retained in the research archive.

October collection ran from October 14, 16:27:50 to October 15, 02:52:00 UTC. Identifiers for 86 September 29 posts were enumerated on October 14, and their bodies fetched on October 16, 03:36--03:40. The records contain no intervening edit, but cannot rule out deletion or unavailability. Both collections use the same roster. Their 199 shared post identifiers include 198 identical saved raw texts and one differing text, giving 26,158 distinct originals across 26,357 observations. Collection-specific inputs remain separate throughout.

The source inventory records 3,174 historical posts and 186 October posts with at least one supplied ancestor; 9,968 and 598 replies, respectively, have no supplied ancestor. A total of 495 later-edited ancestors were excluded (464 historical, 31 October). This is bounded saved context, not a reconstruction of complete conversations. Twenty-four unavailable focal texts remain explicitly marked and do not become empty, complete annotations.

\section{Execution and Repeated Extraction}\label{app:execution}
The first stage comprises 135 development observations and 26,222 production observations in 52 batches. Stages used separate assignments, while agents could process multiple batches in one context. Posts within a batch shared context. Review covered all 1,786 eligible records, changed 226 draft records, and amended relevance-related fields for 134 observations, including 73 relevance-label changes. All 3,033 finalized triples received frozen-category mappings. The primary codebook has seven target categories, twelve aspect categories, and 36 auxiliary aspect nodes.

Every record has explicit processing status. Short response formats were expanded structurally; omitted rows were never defaulted to unrelated. Exact substring checks locate evidence in permitted source fields, preserving multiple matches when necessary. Literal agent outputs, corrections, input hashes, and expansion code are retained. These files do not constitute the provider's full conversation history.

For both checks, sampling excludes development observations and October duplicates of historical post IDs; the main analyses retain collection-specific observations. For the omission check, excluding 8 development observations classified as unrelated in the first stage and 177 October duplicates from the 24,571 initially unrelated observations leaves 24,386. We sampled 360 proportionally across collection--month strata. Fresh agents received no initial labels. All remained unrelated with no candidates or unresolved references, requiring no downstream processing.

For repeated extraction, excluding 127 development observations, 22 October duplicates, and 24 unavailable focal texts from the 1,786 downstream-eligible observations leaves 1,613. These exclusions do not overlap. We drew a disjoint sample of 180: 155 historical and 25 October observations. Sampling was stratified and proportional, with a fixed random seed. Two fresh runs changed input order and batches, followed by separate stance assignments and frozen-category mapping. They did not receive the draft-visible review stage. Comparisons therefore use the first run's pre-review output as baseline. Table~\ref{tab:repeat} reports unweighted agreement; inverse-inclusion-weighted estimates are in the aggregate tables. Sets remove duplicate category combinations and count two empty sets as agreement.

\begin{table}[tp]
\caption{Repeatability on the same 180 observations (percentages). Exact sets contain target category, aspect category, and stance. Jaccard is averaged over posts, assigning one to two empty sets.}
\label{tab:repeat}
\centering
\begin{tabular}{lrrr}
\toprule Comparison & Relevance & Exact set & Jaccard\\
\midrule
Baseline / repeat A & 88.3 & 38.9 & 52.2\\
Baseline / repeat B & 90.6 & 44.4 & 55.7\\
Repeat A / repeat B & 91.7 & 47.2 & 59.2\\
\bottomrule
\end{tabular}
\end{table}

The pre-review baseline and finalized pipeline agree on 95.0\% of relevance labels and 80.0\% of category--stance sets. The latter comparison includes separately performed category mapping as well as source review. Baseline and repeat B category mappings shared a mapping-agent context. Replacing the sampled records with the baseline yields a paired difference of +22.2 points [7.5, 36.2]; repeats A and B give the values in the main text. This experiment changes only the sampled observations.

% Queue the monthly table before the appendix C page boundary.
%\begin{table*}[t]
%\input{tables/monthly_attention.tex}
%\end{table*}
\begin{table}[!htbp]
\small
\setlength{\tabcolsep}{2pt}
% Source: aggregate_data/monthly_attention.csv; historical rows only.
\caption{Historical monthly posts/contributors: retrieved $P,A$, related $R,A_R$. $C_R$: complete related posts; $U$: uncertain relevance; $M$: unavailable text. Related share $100R/P$; allocation upper bound $100(R+U+M)/P$ (not a confidence limit).}
\label{tab:monthly_attention}
\centering
\begin{tabular}{lrrrrrrrrr}
\toprule
Month & $P$ & $A$ & $R$ & $A_R$ & $C_R$ & $U$ & $M$ & Related (\%) & Upper (\%)\\
\midrule
Jan & 2,580 & 139 & 28 & 10 & 25 & 18 & 4 & 1.09 & 1.94\\
Feb & 2,537 & 119 & 62 & 21 & 62 & 15 & 0 & 2.44 & 3.04\\
Mar & 2,739 & 132 & 59 & 15 & 57 & 20 & 2 & 2.15 & 2.96\\
Apr & 2,725 & 138 & 143 & 18 & 142 & 30 & 3 & 5.25 & 6.46\\
May & 2,898 & 138 & 108 & 35 & 102 & 27 & 0 & 3.73 & 4.66\\
Jun & 2,669 & 132 & 81 & 21 & 79 & 17 & 2 & 3.03 & 3.75\\
Jul & 2,820 & 136 & 163 & 32 & 156 & 30 & 4 & 5.78 & 6.99\\
Aug & 2,846 & 125 & 190 & 35 & 188 & 45 & 3 & 6.68 & 8.36\\
Sep & 3,106 & 164 & 357 & 59 & 349 & 110 & 4 & 11.49 & 15.16\\
\bottomrule
\end{tabular}

\end{table}

\section{Aggregation and Composition Details}\label{app:decomposition}
Within each aggregation unit, evaluation, positive-component, and negative-component indicators are logical unions over represented triples. No-expressed-stance and unclear-only units remain in mention counts but not evaluation counts. A positive and a negative evaluation of distinct raw objects in one category cell retain their separate raw states; their cell contains both components. The negative-only sensitivity requires a negative component and no positive component.

Let $r_{at}$ be account $a$'s aspect-mention fraction among complete related posts in month $t$, and $w_{at}$ its weight, normalized across that month's contributors. With $S$ common accounts, $E$ later-only accounts, $X$ earlier-only accounts, and $\bar w_a=(w_{a0}+w_{a1})/2$, $\bar r_a=(r_{a0}+r_{a1})/2$, the exact symmetric decomposition is
\begin{equation}
\begin{split}
R_1-R_0={}&\sum_{a\in S}\bar w_a(r_{a1}-r_{a0})
+\sum_{a\in S}\bar r_a(w_{a1}-w_{a0})\\
&+\sum_{a\in E}w_{a1}r_{a1}
-\sum_{a\in X}w_{a0}r_{a0}.
\end{split}
\end{equation}
For continuing accounts, the first two terms use Kitagawa's symmetric rate-and-composition decomposition~\cite{kitagawa1955}; the last two account for entry and exit. Equal-account weights are reciprocal contributor counts; pooled-post weights are accounts' shares of complete related posts. Even with equal weights, contributor entry changes the weights of continuing accounts. The four terms sum to the observed change before rounding; they do not identify causes of participation.

\section{Evaluation Sensitivities}\label{app:sensitivity}
Figure~\ref{fig:sensitivity} compares the main estimate, the five sensitivity conditions, and the same-post, same-target restriction. The uncertain-relevance sensitivity uses complete resolved triples from related or uncertain posts. The partial-extraction sensitivity retains represented resolved triples from related posts with incomplete extraction; it does not impute missing triples. The tag-only sensitivity excludes posts whose AI connection is supplied only by tags. The unqualified analysis retains triples without conditional, tentative, or other qualifications. Each comparison recomputes post--family units and its own paired-account set.

% Group the appendix comparisons without changing text or table typography.
\begin{table*}[p]
% Sources: aggregate_data/monthly_aspect_mentions.csv; historical July/September rows.
% aggregate_data/aspect_entry_exit_decomposition.csv supplies complete-related contributor counts.
\caption{July--September aspect mentions. $P_a$ counts complete related posts mentioning the aspect. Rel. is their percentage among complete related posts (July: 156; September: 349); Acct. averages within-account mention fractions over all complete-related contributors (31 and 59), including zero mentions. All is the observed mentioning-post percentage among all retrieved posts (2,820 and 3,106). Aspects are nonexclusive.}
\label{tab:july_september}
\centering
\begin{tabular}{lrrrrrrrr}
\toprule
& \multicolumn{4}{c}{July} & \multicolumn{4}{c}{September}\\
\cmidrule(lr){2-5}\cmidrule(lr){6-9}
Aspect & $P_a$ & Rel. (\%) & Acct. (\%) & All (\%) & $P_a$ & Rel. (\%) & Acct. (\%) & All (\%)\\
\midrule
Capability/usefulness & 75 & 48.1 & 50.3 & 2.66 & 99 & 28.4 & 33.2 & 3.19\\
Correctness/technical validation & 26 & 16.7 & 18.4 & 0.92 & 40 & 11.5 & 11.0 & 1.29\\
Understanding/contribution & 20 & 12.8 & 10.5 & 0.71 & 49 & 14.0 & 19.0 & 1.58\\
Usability/work required & 18 & 11.5 & 11.5 & 0.64 & 56 & 16.0 & 17.0 & 1.80\\
Reporting/disclosure/attribution & 26 & 16.7 & 19.4 & 0.92 & 82 & 23.5 & 20.4 & 2.64\\
Review/verification/release & 9 & 5.8 & 6.9 & 0.32 & 21 & 6.0 & 7.1 & 0.68\\
Social/professional/institutional & 41 & 26.3 & 22.4 & 1.45 & 115 & 33.0 & 45.8 & 3.70\\
Costs/environment & 8 & 5.1 & 2.9 & 0.28 & 10 & 2.9 & 4.3 & 0.32\\
\bottomrule
\end{tabular}

\par\bigskip
\noindent\begin{minipage}[c]{\columnwidth}
\centering\includegraphics[width=\linewidth]{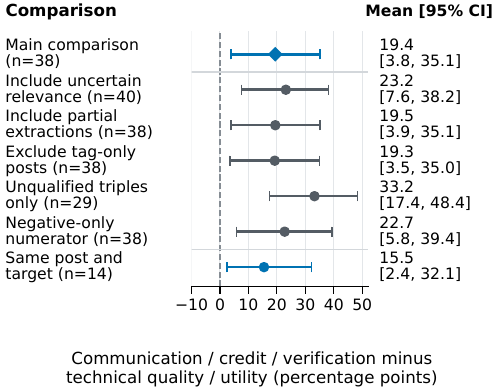}
\captionof{figure}{Procedural-minus-technical negative-component differences across conditions. Points are equal-account means; bars are 95\% paired-account bootstrap intervals; $n$ is the number of paired accounts. The last row averages 27 same-post, same-target differences within 14 accounts. Negative includes mixed except in the negative-only row.}
\Description{A forest plot displays seven comparisons. The primary difference is 19.4 percentage points with an interval from 3.8 to 35.1. All displayed intervals exclude zero.}
\label{fig:sensitivity}
\end{minipage}\hfill
\begin{minipage}[c]{\columnwidth}
% Sources: analysis/tables/rq4_daily.csv and rq4_aspect_daily.csv (primary).
\caption{Daily supports for attention and aspect-expression incidence. $P$: all observed posts; $R$: related posts; $A_R$: related contributors; $C$: complete-related posts; $A_C$: complete-related contributors. Every aspect's incidence uses the account-day subsets of $C$, equally averaging the $A_C$ contributors. October 6 ($\ast$) is included in the post period; October 13 is outside the primary comparison.}
\label{tab:rq4daily}
\centering
\begin{tabular}{lrrrrr}
\toprule Date & $P$ & $R$ & $A_R$ & $C$ & $A_C$\\
\midrule
09-29 & 86 & 6 & 5 & 6 & 5\\
09-30 & 113 & 15 & 7 & 14 & 7\\
10-01 & 87 & 5 & 2 & 5 & 2\\
10-02 & 66 & 3 & 2 & 3 & 2\\
10-03 & 81 & 10 & 2 & 10 & 2\\
10-04 & 58 & 10 & 4 & 10 & 4\\
10-05 & 92 & 7 & 4 & 6 & 4\\
10-06$^{\ast}$ & 115 & 21 & 8 & 21 & 8\\
10-07 & 126 & 18 & 13 & 18 & 13\\
10-08 & 97 & 16 & 10 & 16 & 10\\
10-09 & 130 & 13 & 11 & 12 & 10\\
10-10 & 95 & 16 & 9 & 15 & 9\\
10-11 & 86 & 9 & 4 & 8 & 4\\
10-12 & 108 & 10 & 5 & 9 & 5\\
10-13 & 97 & 8 & 6 & 8 & 6\\
\bottomrule
\end{tabular}

\end{minipage}
\par\bigskip
% Source: analysis/tables/rq4_ols.csv (primary, all; attention and participation only).
\caption{Segmented fits for attention and participation. Window is calendar days per side, including October 6 in the post period; $n_-/n_+$ gives observed dates. Cutoff gaps are percentage points and slopes are percentage points per day. Related-post share uses all observed posts; participation uses the fixed roster of 1,002 accounts.}
\label{tab:rq4fits}
\centering
\begin{tabular}{lrrrrrr}
\toprule Outcome & Window & $n_-/n_+$ & Gap $\tau$ & Pre slope & Post slope & Slope change $\gamma$\\
\midrule
Related-post share & 7 & 7/7 & $+5.31$ & $+0.59$ & $-1.22$ & $-1.81$\\
Roster participation & 7 & 7/7 & $+0.91$ & $-0.03$ & $-0.10$ & $-0.07$\\
Related-post share & 6 & 6/6 & $+5.57$ & $+0.40$ & $-1.08$ & $-1.48$\\
Roster participation & 6 & 6/6 & $+0.88$ & $-0.03$ & $-0.09$ & $-0.06$\\
Related-post share & 5 & 5/5 & $+2.18$ & $+1.64$ & $-0.71$ & $-2.35$\\
Roster participation & 5 & 5/5 & $+0.56$ & $+0.06$ & $0.00$ & $-0.06$\\
\bottomrule
\end{tabular}

\end{table*}

\section{Temporal Comparison and Release Context}\label{app:event}
AGMAI's general release guidelines appeared on September 29, the first pre-period date~\cite{agmai2026sep29}. OpenAI reports consulting AGMAI on release practices~\cite{openai2026mathematicsrelease}.

RQ4 uses only the October snapshot, including the September 29 backfill, without adding overlapping historical observations. Table~\ref{tab:rq4daily} covers all fifteen dates; October 13 is outside the main window. Attention includes all related posts regardless of extraction completeness. Incidence includes complete related posts without identified targets as observations with neither positive nor negative expression.

The sixteen primary incidence fits have seven dates per side. For conditional rates, evaluated dates contain positive, negative, mixed, or explicitly neutral aspect evaluations in complete related posts, regardless of release attribution. No conditional fit meets the five-date-per-side minimum (Table~\ref{tab:rq4aspects}). Undefined rates are not set to zero; daily rates and supports accompany the calculations.

Full output crosses twelve aspects, two components, two estimands, two weighting schemes, three relevance specifications, and three windows (864 fits), plus 288 fits omitting October 6 from September 29--October 12. All 576 incidence fits are computable; none of 576 conditional fits meets the reporting threshold. Table~\ref{tab:rq4aspectsensitivity} summarizes procedural-aspect sensitivities. Incidence alone cannot separate aspect attention, target composition, and expressed stance.

Six of sixteen primary incidence fits predict outside 0--100\% at the cutoff or within the window. Predictions remain unclipped; diagnostics and component counts are retained. Period summaries distinguish pooled-post, weekly equal-account, and mean daily equal-account rates.

\paragraph{Release-specific context.}
A source-only pass classified release references and target scope for 245 eligible October observations using the release title and URL, without stance labels. During October 6--13, 55 posts from 19 accounts directly or contextually refer to the release: 21 related posts from 12 accounts and 34 uncertain posts from 14 accounts, with overlapping account sets. Complete related extraction leaves 15 posts with release-specific targets from 11 accounts, all present in the historical corpus. Table~\ref{tab:event} summarizes this descriptive subset, which does not select the temporal-analysis corpus.

% Keep only the wide tables in this float; explanatory prose uses the normal columns.
\begin{table*}[p]
% Sources: analysis/tables/rq4_aspect_main_report.csv and rq4_aspect_ols.csv (primary, seven-day window).
\caption{Aspect-specific expressions in September 29--October 5 and October 6--12. Count pairs are pre/post numbers of posts, not percentages; a post can contribute to both polarities. The pre/post denominators contain 54/99 complete-related posts. The incidence gaps $\tau_+$ and $\tau_-$ are in percentage points: within each account-day, the numerator is posts expressing the indicated polarity on that aspect and the denominator is all complete-related posts, including zero aspect mentions, followed by equal averaging across accounts. All 16 incidence fits use seven dates on each side. The final column gives dates with evaluated posts for conditional stance; all conditional fits fail the five-dates-per-side gate. $\dagger$: at least one fitted cutoff or in-window value lies outside $[0,100]\%$; predictions are not clipped.}
\label{tab:rq4aspects}
\centering
\begin{tabular*}{\textwidth}{@{\extracolsep{\fill}}lrrrrrr@{}}
\toprule Aspect & \shortstack{Evaluated posts\\pre/post} & \shortstack{Positive posts\\pre/post} & \shortstack{Negative posts\\pre/post} & $\tau_+$ & $\tau_-$ & \shortstack{Conditional\\pre/post dates}\\
\midrule
Capability/usefulness & 1/15 & 1/9 & 0/7 & $+12.79^{\dagger}$ & $+4.40$ & 1/7\\
Correctness/technical validation & 2/6 & 1/1 & 1/5 & $-2.61^{\dagger}$ & $+2.26^{\dagger}$ & 2/3\\
Understanding/contribution & 2/5 & 1/4 & 1/2 & $+3.84^{\dagger}$ & $-9.88^{\dagger}$ & 2/4\\
Usability/work required & 4/8 & 0/3 & 4/5 & $+2.99$ & $-0.69$ & 4/5\\
Reporting/disclosure/attribution & 8/10 & 2/1 & 6/9 & $-4.63$ & $+13.20$ & 4/6\\
Review/verification/release & 0/4 & 0/2 & 0/2 & $+4.33$ & $+4.18$ & 0/3\\
Social/professional/institutional & 5/20 & 3/14 & 3/10 & $+3.88^{\dagger}$ & $+1.07$ & 3/7\\
Costs/environment & 0/5 & 0/2 & 0/3 & $+1.03$ & $+0.36$ & 0/3\\
\bottomrule
\end{tabular*}

\par\medskip
% Sources: analysis/tables/rq4_aspect_ols.csv and rq4_aspect_eventday_sensitivity.csv (incidence; communication and review).
\caption{Sensitivity of expression-incidence gaps for the two procedural aspects (percentage points). The first three columns vary days per side. The next three retain seven days per side while changing relevance or weighting. The final column omits October 6 from September 29--October 12 (seven pre and six post dates); other columns include it. Rates are equal-account except in the pooled-post column; all fits weight dates equally. $\dagger$: a fitted cutoff or in-window value outside $[0,100]\%$.}
\label{tab:rq4aspectsensitivity}
\centering
\begin{tabular*}{\textwidth}{@{\extracolsep{\fill}}lrrrrrrr@{}}
\toprule Aspect / polarity & 7 days & 6 days & 5 days & \shortstack{Include\\uncertain} & \shortstack{Exclude\\tag-only} & \shortstack{Pooled\\posts} & \shortstack{Omit\\Oct. 6}\\
\midrule
Reporting/disclosure/attribution / positive & $-4.63$ & $+1.75^{\dagger}$ & $+5.00^{\dagger}$ & $-2.45$ & $-5.10$ & $-2.55$ & $-4.21$\\
Reporting/disclosure/attribution / negative & $+13.20$ & $+10.52$ & $-3.29^{\dagger}$ & $+18.24^{\dagger}$ & $-1.93$ & $+6.50$ & $+6.86$\\
Review/verification/release / positive & $+4.33$ & $+4.23$ & $+3.75$ & $+3.68$ & $+5.30^{\dagger}$ & $+3.40$ & $+2.67$\\
Review/verification/release / negative & $+4.18$ & $+3.88$ & $+3.08$ & $+3.33$ & $+4.40$ & $+3.17$ & $+7.79^{\dagger}$\\
\bottomrule
\end{tabular*}

\par\medskip
% Sources: aggregate_data/event_openai_mathematics_release_2026_10_06_family_summary.csv;
% aggregate_data/event_including_uncertain_relevance_family_summary.csv;
% aggregate_data/event_summary.csv; aggregate_data/event_including_uncertain_relevance_paired_summary.csv;
% aggregate_data/event_period_family_summary.csv (October 6 note).
\caption{Release-specific coverage of complete related posts, October 6--13; sensitivity adds complete uncertain-relevance posts. Mention/evaluation cells give posts/accounts (P/A) with in-scope family targets; evaluations require positive, negative, explicitly neutral, or mixed stance. $N$ counts evaluated posts with a negative component. Negative (\%) pairs give percentages pooled over evaluated posts/averaged equally over evaluating accounts. Dashes indicate zero evaluation denominators; rows can overlap. In the main case, October 6 contributes one technical mention and no evaluation; all evaluations occur on October 7--13.}
\label{tab:event}
\centering
\begin{tabular*}{\textwidth}{@{\extracolsep{\fill}}lrrrrrrrr@{}}
\toprule
& \multicolumn{4}{c}{Main (15 posts, 11 accounts)} & \multicolumn{4}{c}{Sensitivity (30 posts, 14 accounts)}\\
\cmidrule(lr){2-5}\cmidrule(lr){6-9}
Aspect family & \shortstack{Mention\\P/A} & \shortstack{Evaluation\\P/A} & $N$ & \shortstack{Neg. (\%)\\Pooled/eq.} & \shortstack{Mention\\P/A} & \shortstack{Evaluation\\P/A} & $N$ & \shortstack{Neg. (\%)\\Pooled/eq.}\\
\midrule
Technical quality/utility & 8/7 & 5/5 & 3 & 60/60 & 19/9 & 8/7 & 5 & 62.5/64.3\\
Communication/credit/verification & 7/6 & 2/2 & 2 & 100/100 & 10/7 & 2/2 & 2 & 100/100\\
Other aspects & 5/5 & 3/3 & 2 & 66.7/66.7 & 9/7 & 3/3 & 2 & 66.7/66.7\\
Unspecified/general/unresolved & 2/2 & 0/0 & 0 & ---/--- & 5/5 & 0/0 & 0 & ---/---\\
\bottomrule
\end{tabular*}

\end{table*}

\end{document}